\documentclass[prl,showpacs,twocolumn,amsmath,amssymb,superscriptaddress]{revtex4-1}

\usepackage{graphicx}% Include figure files
\usepackage{dcolumn}% Align table columns on decimal point
\usepackage{bm}% bold math

\newcommand{\anferm}{\psi^{\phantom{\dagger}}}
\newcommand{\crferm}{\psi^{\dagger}}

\begin{document}

\title{Macroscopic Degeneracy and Emergent Frustration in a Honeycomb Lattice Magnet}

\author{J\"orn W.F. Venderbos}
\email{	j.venderbos@ifw-dresden.de }
%\affiliation{IFW Dresden, P.O. Box 27 01 16, D-01171 Dresden, Germany}

\author{Maria Daghofer}
%\affiliation{IFW Dresden, P.O. Box 27 01 16, D-01171 Dresden, Germany}

\author{Jeroen van den Brink}
\affiliation{IFW Dresden, P.O. Box 27 01 16, D-01171 Dresden, Germany}

\author{Sanjeev Kumar}
\affiliation{Indian Institute of Science Education and Research (IISER) Mohali, Knowledge city, Sector 81, Mohali 140306, India
}

\date{\today}

\begin{abstract}
Using a hybrid method based on fermionic diagonalization and classical Monte Carlo, we investigate the interplay between itinerant and localized spins, with competing double-
 and super-exchange interactions, on a honeycomb lattice. For 
moderate superexchange, 
a geometrically frustrated triangular lattice of hexagons forms spontaneously. For slightly larger superexchange a dimerized
groundstate is stable that has macroscopic degeneracy. The presence of these states on a non-frustrated honeycomb lattice 
highlights novel phenomena in this itinerant electron system: emergent geometrical frustration and degeneracy related to a symmetry intermediate between local an global.

\end{abstract}

\maketitle

The Kondo lattice model (KLM) is probably the most celebrated starting point for the investigation of the interplay between localized spins and itinerant electrons \cite{KLM_review}. It provides the canonical explanation for the Kondo effect and for the heavy-fermion behaviour observed in many  materials \cite{heavy-fermions}. 
Motivated by the search for topologically non-trivial states of matter, several groups have recently studied the itinerant KLM on frustrated lattices, such as the triangular or the pyrochlore one, and have shown that due to the strong geometrical frustration scalar-chiral types of magnetic ordering emerge \cite{Shindou01,triangular,Chern}. 

The physics of the KLM on non-frustrated lattices, such as the square
and cubic one, has been studied extensively. In particular the limit
of strong coupling and large localized moments, where  the KLM goes
over into the double-exchange (DE) model, is directly relevant to the
colossal magnetoresistance effect in perovskite manganites
\cite{Dagotto_book,Brey_PRL-Dagotto_PRL,SK-PM}. In such cases, the
competition between DE and antiferromagnetic (AFM) superexchange can
lead  to canted spin states or phase separation~\cite{Dagotto_book}.  
Although the honeycomb lattice is also bi-partite, it has the smallest possible coordination number for proper 2D
lattices. 
That the honeycomb lattice can support physical phenomena
fundamentally different from square lattices, is illustrated by recent
Quantum Monte Carlo calculations~\cite{Meng10},  
which identify a novel spin-liquid phase for the Hubbard model on the honeycomb
lattice, a finding supported by analytical
studies~\cite{Wang10,Lu10,Vaezi10}.

In this Letter, we investigate the consequences of the competition between AFM superexchange and
ferromagnetic (FM) DE on the honeycomb lattice.  
We find that two exotic ground states exist between the 
trivial,
fully FM and AFM phases. In the first, nearer
to the FM state, the spins self-organize into FM hexagons
that are coupled antiferromagnetically. Since the 
hexagonal
rings form a
\emph{frustrated} triangular lattice, their order is reminiscent of
the Yafet-Kittel state~\cite{YafetKittel}. The competition between 
isotropic 
magnetic
interactions thus causes geometric frustration to emerge in a
non-frustrated lattice.  

For slightly stronger AFM interactions, we find the 
\emph{exact}
groundstate to
consist of independent FM dimers containing one electron each. Apart
from the requirement that the alignment of adjacent dimers be AFM, they
are independent. The groundstate of this $N$-spin system therefore has a high
degeneracy $\propto 2^{\sqrt{N}}$. While the macroscopic degeneracy
$\propto e^{\alpha N}$ in (spin) ice is caused by the local symmetry of the
frustrated tetrahedra~\cite{Castelnovo08,Gardiner10}, our $\sqrt{N}$ exponent 
indicates the presence of
an `intermediate' symmetry -- a symmetry between local and
global~\cite{Nussinov:2009p2519}. 
It is remarkable that
this highly degenerate
groundstate manifold arises as an \emph{emergent} effect in a
Hamiltonian that itself does not have such a symmetry.

In many
materials,  
the essence of the
electronic structure is captured by interacting spins and electrons on
a honeycomb lattice. 
Interactions between impurity
magnetic moments on the honeycomb lattice of graphene
have been intensely studied in a Ruderman-Kittel-Kasuya-Yosida (RKKY)
framework~\cite{RKKY_graphene} and using the KLM~\cite{KLM_graphene}. Going
beyond RKKY is even more important in transition metal oxides, 
e.g., 
Bi$_3$Mn$_4$O$_{12}$(NO$_3$)~\cite{BiMnONO} 
or Li$_2$MnO$_3$\cite{LiNiMnO}, 
with Mn ions %residing 
on a honeycomb
lattice. 

The Hamiltonian corresponding to the one-band DE model in the presence of competing AFM superexchange interactions on a honeycomb lattice is
\begin{equation} \label{eq:ham}
H = -\sum_{\langle ij\rangle} \, t_{ij} (\crferm_i\anferm_j + H.c.) 
+ J_{\textrm AF}\sum_{\langle ij\rangle} {\bf S}_i \cdot  {\bf S}_j ,
\end{equation}
where $\crferm_i$ and $\anferm_i$ are the fermionic creation and
annihilation operators, respectively. In accordance with the DE scheme
these fermions have their spin 
aligned with the on-site spins ${\bf S}_i$. The on-site core spins are
treated as classical spins with $|{\bf S}_i| =1 $ and thus can be
specified by their polar and azimuthal angles
$(\theta_i,\phi_i)$. Both sums are over nearest neighbors.
Due to the alignment of electron spin to the
core spins, the hopping amplitude depends on the direction of the core
spins, %specifically 
$t_{ij} =
t_0[\cos(\theta_i/2)\cos(\theta_j/2)+\sin(\theta_i/2)\sin(\theta_j/2)e^{-i(\phi_i-\phi_j)}
]$ \cite{Dagotto_book}. The strength of the AFM super-exchange is
given by $J_{\textrm AF}$ and all energies are in units of the hopping
amplitude $t_0$. 
To guarantee an unbiased search for
groundstate candidates, we employ a well-established hybrid method of
exact diagonalization (ED) for the bilinear fermionic part of the
Hamiltonian and Monte Carlo (MC) for the classical spins
\cite{Dagotto_book}. 
Each MC configuration is defined by
a given core spin texture and Markov chains are generated by
diagonalizing the fermionic problem 
for each configuration update. 
We also 
make use of the travelling cluster
approximation (TCA), which 
has proven its validity and success in earlier 
studies on a similar class of models~\cite{SK-PM}, 
to go to
larger lattice sizes.
%\cite{TCA}. 
We report here results based on
calculations on an $N=12^2$ honeycomb lattice, using a cluster of size
$N_c = 6^2$. In the MC routine we use $\sim10^4$ steps for
equilibration and the same number of steps for thermal averaging. 
%In this work we 
We focus on the case of a half-filled band, which refers to 1/2 an
electron per site, equivalent 
to quarter-filling in the spinful problem. For selected parameter
values, the MC procedure was  
further refined
by an optimization routine  
that diminishes
thermal fluctuations~\cite{Yu:Prb09}.

\begin{figure}
\center 
\includegraphics[
width=.8\columnwidth]{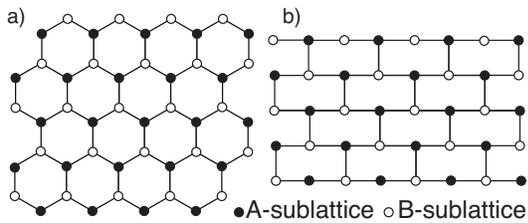} 
\caption{\label{fig:lattice} 
Schematic view of (a) the honeycomb lattice and (b) the brick-wall
lattice having the same topology.%
}
\end{figure}

To identify the magnetically ordered states, we calculate the spin structure factor $S({\bf q}) = \frac{1}{4N^2}\sum_{i,j}\, \langle {\bf S}_i\cdot {\bf S}_j \rangle e^{i{\bf q}\cdot ({\bf r}_i-{\bf r}_j)} $, 
where $\langle \ldots \rangle$ is a thermal average and ${\bf r}_i$ is the position space vector of site $i$. For a clear understanding of the real-space structure of the magnetic states it is helpful to look at $S({\bf q})$ on a square geometry [see Fig.~\ref{fig:lattice}(b)].
A specific long-range ordering is expressed as the point in the
Brillouin zone where the structure factor shows a peak.
To analyze the electronic properties we compute the
density of states (DOS) as $D(\omega) = \langle \tfrac{1}{N} \sum_k
\delta(\omega-\epsilon_k) \rangle$ and approximate the delta-function
by a Lorentzian with broadening $\gamma$.

\begin{figure}
\center \includegraphics[
width=0.9\columnwidth]{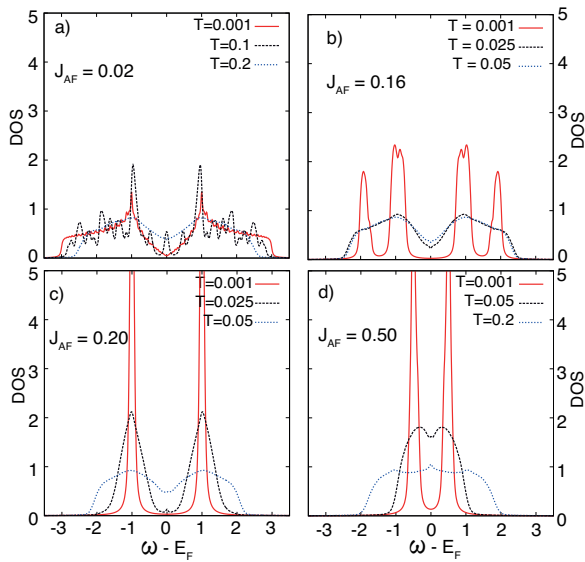} \caption{
\label{fig:DOS} (Color online) (a)-(d) DOS
at low,
intermediate and high temperatures for different values of $J_{\textrm
  AF}$ ($\gamma=0.04$). In (a) the $T=0.001$ curve shows the DOS of free fermions on
a honeycomb lattice in the thermodynamic limit. In (d) the $T=0.001$
curve represents a gapped insulating phase, the seemingly finite DOS
at $E_F$ being a broadening effect.  
%MD: Moved that here:
Except for the FM phase, all ground states are gapped.
}
\end{figure}

\begin{figure}
\center \includegraphics[
width=0.8\columnwidth]{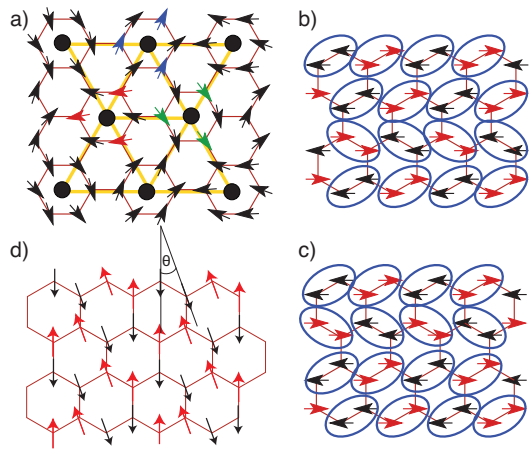} 
\caption{\label{fig:spinstates} 
(Color online) (a) Snapshot from MC simulations supplemented by optimization routines showing an emergent triangular lattice (black circles) formed by FM hexagons at
$J_{\textrm AF}=0.14$. Spins within
each hexagon are almost FM, a small canting angle between
groups of three is illustrated by shading.
The colored spins illustrate the $2\pi/3$-angle order of the
Yafet-Kittel state.
Schematic view of, (b)-(c) two dimer states related by a translational symmetry (see text), and (d) a canted dimer state.
}
\end{figure}

In the absence of super-exchange interaction ($J_{\textrm AF}=0$), the
spins order ferromagnetically, as expected from the DE mechanism. The
fermionic problem is then equivalent to non-interacting spinless
electrons on a honeycomb lattice, giving rise to a dispersion and DOS
that is well-known from graphene [see
Fig. \ref{fig:DOS}(a)]. Introducing a small $J_{\textrm AF}$ still
leads to a FM ground state. 
At $J_{\textrm AF}\approx 0.14$, the FM state becomes unstable and
gives way to a state with $S({\bf q})$ peaked at $\tfrac{2}{3}(\pi,0)$
(and the points related to it by symmetry).
and with the peculiar four-peak DOS  shown in Fig.~\ref{fig:DOS}(b).  
Real-space snapshots show that a superlattice formed of hexagons
emerges at low temperatures $T$, as depicted in Fig.~\ref{fig:spinstates}(a). 
This result was corroborated by zero-temperature optimization of the
spin pattern. Spins within one hexagon are almost FM, the allowed
energies for electrons moving on a six-site ring are $-2t_0\cos
0=-2t_0$ and $- 2t_0\cos \pi/3 = - t_0$, with twice as many states at
$-t_0$, which gives precisely the DOS seen in
Fig.~\ref{fig:DOS}(b). Coupling between the hexagons is AF, but since
they occupy a frustrated triangular lattice, see
Fig.~\ref{fig:spinstates}(a), perfect AFM order is not possible. The
hexagons instead are at an angle of $\approx 2\pi/3$, corresponding to
the Yafet-Kittel state~\cite{YafetKittel} well known for the
triangular lattice, leading to the signals at $\tfrac{2}{3}(\pi,0)$ in
$S({\bf q})$.  
Thus a geometrically frustrated triangular lattice emerges spontaneously from 
isotropic, competing interactions on
the non-frustrated honeycomb lattice.

For $0.18 \leq J_{\textrm AF} \lesssim 0.25$, 
we find a state consisting of classical dimers.
The dimers each consist 
of two spins aligned in parallel, 
they cover the lattice 
in such a way that the neighboring dimers are anti-parallel with
respect to each other. In Fig.~\ref{fig:spinstates}(b) and
\ref{fig:spinstates}(c) we show two possible dimer configurations. In
this spin texture, 
the electron kinetic energy
reduces to that of uncoupled two-level problems, having only two
eigenenergies $\pm t_0$. The DOS is therefore given by $D(\omega) =
\delta(\omega-t_0)/2+\delta(\omega+t_0)/2$, in excellent agreement with MC
calculations [see Fig. \ref{fig:DOS}(c)]. The dimer state  can be
understood as a trade-off between the FM ordering and the AFM
ordering: the electrons are allowed to populate all the $-t_0$ levels
(which is more 
favorable compared to AFM) and the spins are
anti-parallel with respect to two of their nearest neighbors  
 (which is more favorable compared to FM).  

Interestingly, the dimer ground state of this quantum system has a
macroscopic degeneracy, i.e., there is a macroscopically large number
of ways to cover the lattice by dimers such that the neighboring
dimers are anti-parallel.  
One %simple 
way to see the degeneracy is to start covering lattice rows % of
in Fig.~\ref{fig:lattice}(b) by dimers. It is easy to see
that having fixed the dimer pattern in the 1st row, there are two
independent ways of covering each subsequent row, giving 
$2^{\sqrt{N}-1}$ states for a $N$-site lattice.
The fact that there is thus no long-range order along the $y$
direction of the brick-wall is reflected in $S({\bf q})$, which
becomes finite along \emph{lines} in momentum space, as in
compass models~\cite{Mishra04,Biskup05,Nussinov04,Compass_Heisenberg}.
In the 2D compass model, different degenerate configurations can be
reached by flipping a row of spins. The corresponding $\sqrt{N}$
operators commute with the compass Hamiltonian and thus define an
\emph{intermediate} symmetry, i.e., between a local, gauge-like 
($\propto e^N$) symmetry and global one 
(independent of $N$)~\cite{Nussinov:2009p2519}. The magnetic order
parameter that obeys the intermediate symmetry is consequently of
nematic type. 
In the dimer state, the minimal symmetry operations 
%instead 
involves translation
of all spins in two adjacent zig-zag rows by one lattice spacing, $\sigma_{ij} \mapsto \sigma_{ij+1}$ [$\sigma_{ij}$
is the spin at site $(i,j)$]. An example for two dimer configurations connected
by such an operation is given in Figs.~\ref{fig:spinstates}(b) and
\ref{fig:spinstates}(c), where the second and third rows were
shifted. However, this operator does \emph{not commute with the Hamiltonian}
Eq.~(\ref{eq:ham}), and the intermediate symmetry is thus rather a property
that \emph{emerges} in the system's ground state, similar to the case
of striped phases at fractional filling in the regime of narrow
bandwidth and small  Jahn-Teller coupling in a model used for
manganites~\cite{Liang2011}. This intermediate macroscopic degeneracy
should lead to a
large specific heat at low temperature. 

For strong super-exchange coupling, there is a continuous way in which
the dimer state can approach the AFM ordered state, captured by a
canting angle $\theta$ [see Fig. \ref{fig:spinstates}(d)], which is
the angle between the two spins forming a dimer in the pure dimer
phase. The spins remain antiparallel to those of the neighboring
canted dimers. In this way, the two-level dimer systems remain
uncoupled. The hopping amplitude between the two spins in the dimer is
renormalized by the DE mechanism to $t_0
\cos(\theta/2)$. The DOS for a canted dimer state is consequently
given by $D(\omega) = \delta(\omega-t_0
\cos(\theta/2))/2+\delta(\omega+t_0 \cos(\theta/2))/2$, as can indeed
be observed in the %The %calculated 
DOS for $J_{\textrm AF}=0.50$ shown in Fig.~\ref{fig:DOS}(d).
The canted-dimer groundstate has a gap $\propto \cos(\theta/2)$ at the chemical potential,
which shrinks as $\theta$ approaches $\pi$ for
$J_{\textrm{AF}}\to\infty$. At finite $T$, the two peaks widen and
merge due to thermal spin fluctuations, leading to a metal with
reduced band width, see Fig.~\ref{fig:DOS}(d).  
This canted state retains the macroscopic
degeneracy inherent to the AFM dimer state discussed 
above
-- also this ordering is therefore of nematic type.  

\begin{figure}
\center \includegraphics[width=\columnwidth]{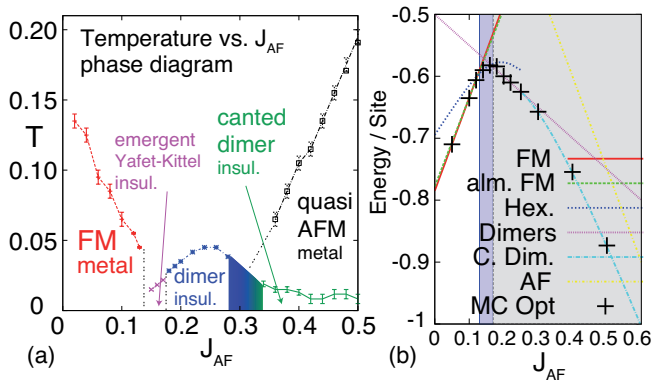} \caption{\label{fig:T_J}  
(Color online) (a) $T$-$J_{\textrm AF}$ phase diagram at half-filling obtained by MC for a
$12\times 12$ lattice.  
(b) The energy of the various states:
``alm. FM'' refers to a spiral with the longest wavelength
supported by the lattice, converging  to FM in the
thermodynamic limit.  Similar finite-size effects are reported in
doped 1D and 2D lattices~\cite{Reira97}. ``Hex'' denotes the emergent
Yafet-Kittel order between hexagons depicted in
Fig.~\ref{fig:spinstates}(a), the energy was optimized with respect to
the canting angle within the hexagons. ``Dimers'' and ``C. Dim'' are
the highly degenerate FM and canted dimer states, and ``AFM'' denotes
perfectly AFM order. The black crosses are energies obtained by
unbiased MC and a subsequent energy optimization. 
}
\label{fig:fig4}
\end{figure}

Our results are in good agreement with elementary energy
considerations. The energy per site varies as $3J_{\textrm AF}/2$ and
$-J_{\textrm AF}/2$ for the FM and the dimer states,
respectively. This would imply a phase transition at $J_{\textrm AF}
\approx 0.15$, the FM state is indeed stable for $J_{\textrm AF} \lesssim
0.14$ and the dimers for $J_{\textrm AF} \gtrsim 0.18$. In between,
the emergent Yafet-Kittel state, with a more complex energy
dependence, is favorable, see Fig.~\ref{fig:fig4}(b). 
The energy per site for the canted dimer state is $-(2J_{\textrm AF}-\cos(\theta)J_{\textrm AF}+t_0\cos(\theta/2) )/2$.
By differentiating with respect to the canting angle $\theta$, one easily
obtains that canting becomes favorable for $J_{\textrm AF}\geq 0.25$ and
that the optimal energy is then given by $-3J_{\textrm AF}/2 - t_0^2/(16J_{\textrm AF})$. This
is reflected in the behavior of the ordering temperature for the dimer
state, which starts decreasing at $J_{\textrm AF}=0.25$ [see
Fig. \ref{fig:T_J}(a)]. 

The results are summarized in Fig.~\ref{fig:T_J}. In the finite-$T$
phase diagram Fig.~\ref{fig:T_J}(a), phase boundaries for the FM and quasi-AFM
regions are obtained by determining the inflection point in the
$\langle M \rangle (T)$ and $\langle \overline{M} \rangle (T)$
($\overline{M}$ denotes staggered magnetization) curves. The onset of
dimer and other phases is determined by tracking the temperature
dependence of the spin structure factor and the characteristic
features in the DOS. 
Figure~\ref{fig:T_J}(b) compares the ground-state energies of the
various phases and perfectly agrees with the unbiased numerical data,
indicating that we have identified the ground states correctly. 

In a full quantum treatment of the spin system additional quantum fluctuations can affect the stability of these ordered phases. Here one anticipates 
the FM building blocks (hexagons and dimers) to be robust as they are stabilized by a
substantial DE energy, and the FM state remains an eigenstate of the
hexagon (dimer) for quantum spins. The Yafet-Kittel ordering between the large
total spins of the hexagons is expected to be more classical, and thus
robust, than for $S=1/2$, where it is found for $T\to
0$~\cite{fluct}. If one can describe the magnetism of a dimer state by an effective NN AFM Heisenberg model, then this model remains the same if one performs the
operation illustrated in Fig.~\ref{fig:spinstates}. The emergent symmetry would thus commute with the effective low-energy Hamiltonian so that the corresponding degeneracies are preserved.

We conclude that the 
isotropic
double-exchange model with competing super-exchange interactions on the {\it non-frustrated} honeycomb lattice has 
an unexpectedly 
rich phase diagram with exotic magnetic phases. 
In one of these, FM rings become the essential building blocks, which
form a frustrated triangular lattice and are antiferromagnetically
coupled. The stabilization of such frustrated spin states on a
bipartite honeycomb lattice, without explicit frustration, is  
so far unique and an example 
of geometrical frustration emerging from competing interactions.
Another novel phase consists of FM dimers ordered antiferromagnetically
and has a $2^{\sqrt{N}}$ degeneracy. This is reminiscent of compass
models, but in the present case the
corresponding symmetry is not a property of the Hamiltonian given a
priori, but rather a property that emerges in the system’s ground
state~\cite{Mishra04,Biskup05,Nussinov04,Compass_Heisenberg,Liang2011}.
 These phenomena are not only relevant in a theoretical context, 
immediately raising the question which other models share such
features and how further residual interactions might affect the degeneracy, but
pertains 
in particular to honeycomb manganese oxides, which form a promising
class of materials to realize these novel types of highly frustrated
states harboring macroscopic degeneracies. 

This research was supported by the Interphase Program of the Dutch Science Foundation NWO/FOM and by the Emmy-Noether program of the DFG.

{}

\end{document}